\documentclass[%
 aip,
 jmp,%
 amsmath,amssymb,
 reprint,%
 unsortedaddress
]{revtex4-1}

\usepackage{graphicx}
\usepackage{dcolumn}
\usepackage{bm}
\usepackage{multirow}

\usepackage{mathtools}
\usepackage{color}

\newcommand{\Uvec}{\boldsymbol{U}}
\newcommand{\Rvec}{\boldsymbol{R}}
\newcommand{\Pvec}{\boldsymbol{P}}
\newcommand{\Qvec}{\boldsymbol{Q}}

\newcommand{\Lcalvec}{\boldsymbol{\mathcal{L}}}
\newcommand{\Ocalvec}{\boldsymbol{\mathcal{O}}}
\newcommand{\Dcalvec}{\boldsymbol{\mathcal{D}}}

\newcommand{\Pmat}{\mathbf{P}}
\newcommand{\Qmat}{\mathbf{Q}}
\newcommand{\Dmat}{\mathbf{D}}
\newcommand{\Nmat}{\mathbf{N}}

\newcommand{\Nbas}{\mathcal{N}}
\newcommand{\Curve}{\boldsymbol{\mathcal{C}}}
\newcommand{\CurveOneDim}{\mathcal{C}}
\newcommand{\transp}{\text{T}}

\begin{document}
\preprint{AIP/123-QED}
\title[]{Integrated Reaction Path Processing from Sampled Structure Sequences}
\author{Michael A. Heuer}
\author{Alain C. Vaucher}
\author{Moritz P. Haag}
\author{Markus Reiher}
\email[Corresponding author: ]{markus.reiher@phys.chem.ethz.ch}
\affiliation{
ETH Z\"urich, Laboratorium f\"ur Physikalische Chemie, Vladimir-Prelog-Weg 2, CH-8093 Z\"urich, Switzerland
}
\date{\today}

\begin{abstract}
Sampled structure sequences obtained, for instance, from real-time reactivity explorations or first-principles molecular dynamics simulations contain valuable information about chemical reactivity.
Eventually, such sequences allow for the construction of reaction networks that are required for the kinetic analysis of chemical systems.
For this purpose, however, the sampled information must be processed to obtain stable chemical structures and associated transition states.
The manual extraction of valuable information from such reaction paths is straightforward but unfeasible for large and complex reaction networks.
For real-time quantum chemistry, this implies automatization of the extraction and relaxation process while maintaining immersion in the virtual chemical environment.
Here, we describe an efficient path processing scheme for the on-the-fly construction of an exploration network by approximating the explored paths as continuous basis-spline curves.
\end{abstract}

\maketitle
\setlength{\parindent}{0cm}
\setlength{\parskip}{0.6em plus0.2em minus0.1em}

\section{Introduction}

A large number of chemical systems highly relevant to industrial chemistry,\cite{patil2012a,vinu2012a} biology,\cite{ross2008a} and environmental chemistry\cite{vereecken2015a} are still poorly understood due to the complexity of their reaction networks.\cite{helfferich2004a,vinu2012a}
Understanding their reactivity requires the knowledge of relevant stable intermediates and of the energetics of their interconnecting elementary reactions.
This knowledge allows for calculating reaction rate constants and conducting kinetic analyses to identify favored reaction routes and bottlenecks.

With increasing system size, the number of stable compounds, alternative routes, and side reactions becomes increasingly high.
Accordingly, it is an unfeasible task to identify all possible routes manually.
Options out of this dilemma are computer-driven explorations based on heuristic concepts, exploratory dynamics, and interactively steered explorations.

First-principles (\textit{ab initio}) molecular dynamics (AIMD),\cite{marx2009} and reactive molecular dynamics,\cite{doentgen2015a,saitta2014a,wang2014a} record the evolution of a system on one adiabatic potential energy surface (PES).
A major challenge in AIMD is to escape deep wells on the PES which otherwise leads to long simulation times.
Different strategies exist to increase the frequency of such rare events.\cite{huber1994,laio2002,iannuzzi2003,wang2014a,stone2001,stone2010b}
Heuristic models represent a second approach for the study of complex chemical reactivity,\cite{zimmerman2013a,zimmerman2015a,rappoport2014a,bergeler2015a,simm2017a} where intermediates and chemical reactions are generated from a set of heuristic rules based on chemical concepts or quantum mechanical descriptors.

A third approach, which may be considered complementary to the other two or on its own, is real-time quantum chemistry.\cite{marti2009,haag2011,bosson2012,haag2013,haag2014a,haag2014b,vaucher2016a,muehlbach2016a,vaucher2016b}
Real-time quantum chemistry immerses chemists into the reactivity exploration process through interactivity and specialized hardware.
Fast quantum chemical methods deliver the electronic structure and nuclear gradients of some target chemical system in real time.
This allows chemists to interact with molecules and thereby induce and screen chemical reactions.
For this purpose, chemists can manipulate the structure of a molecular system with the computer mouse or with a force-feedback haptic device, which also renders the quantum-chemical force on an atom resulting from the manipulation.

For all three approaches, the generated structures and reaction paths must be processed to deliver optimized intermediates and elementary reactions.
Often, the information of interest can be summarized in a set of stable chemical structures and a set of minimum energy paths connecting them.
A minimum energy path (MEP) is connecting a transition state with two minimum structures by following the direction of steepest descent.
Since there can be multiple transition states connecting two given structures, there can also be multiple MEPs between those structures.
One may distinguish between local MEPs and the global MEP characterized by the transition state with the lowest energy.

Both, AIMD and real-time quantum chemistry, generate sequences of sampled structures and associated total electronic energies.
These sampled sequences contain valuable information about reaction events.
However, they often contain noise in the form of atomic or molecular motions that are hardly relevant for reactivity analysis and should be filtered out.

A reaction network emerging from an exploration can be depicted as a graph, in which nodes represent single molecular structures, usually stable structures on the potential energy surface, and edges represent the paths connecting them.
Reaction networks are a compact representation of reactivity information, and can be analyzed with standard graph algorithms.\cite{even2011a}

In this work, we propose an integrated process for reaction network construction from sequences of structures and associated energies.
We consider a $(3N+1)$-dimensional space including a dimension for the energy along with $3N$ dimensions for the $N$ Cartesian nuclear coordinates.
The exploration path is approximated by a curve in this space represented by a cubic basis-spline (B-spline) curve.
The resulting continuous parametric curve enables fast processing, segmentation, and construction of an exploration network.
To obtain the energetics of the reaction, the exploration network is then relaxed. This yields the stationary points and MEPs interconnecting them on the PES.

This strategy is related to approaches that filter out reaction events from long molecular dynamics simulations.
For instance, the so-called nebterpolation method analyzes molecular dynamics simulations to identify reaction events, which are then processed to filter out minimum energy paths.\cite{wang2016a}
Another work introduces a new type of reaction coordinate to study complex dynamics simulations and filter through statistical noise.\cite{mcgibbon2017a}
A conceptual difference to our approach is that such protocols deliver a series of distinct events extracted from the simulation, while we aim to provide a comprehensible representation of the full reactivity exploration.

\begin{figure}[htb]
\includegraphics[scale = 0.5]{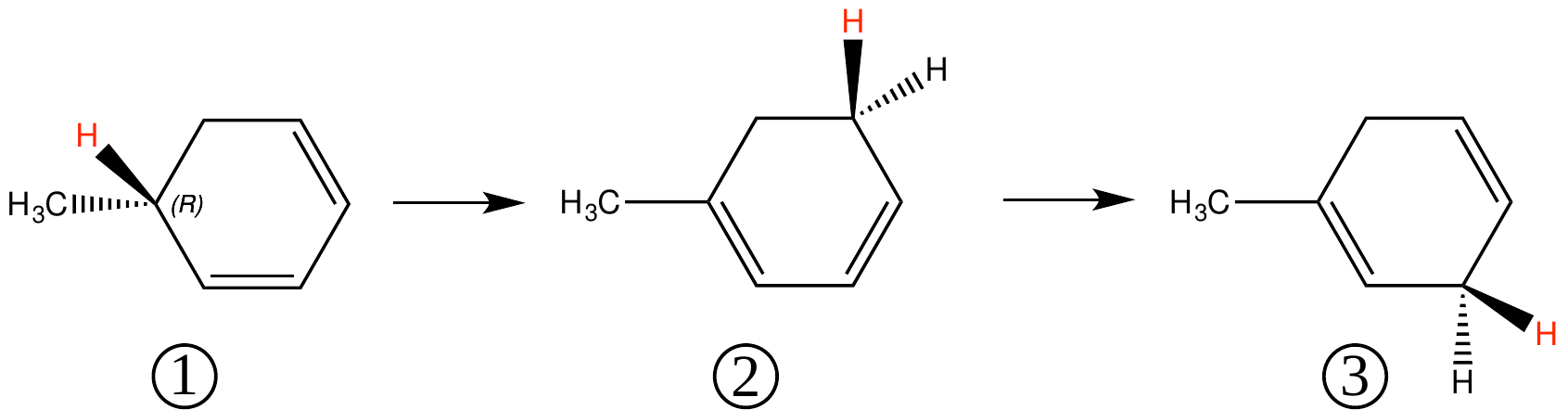}
\caption{Lewis structures of the stable intermediates of the interactively explored hydrogen shift reactions of
\textcircled{\scriptsize{1}} (\textit{R})-5-methylcyclohexa-1,3-diene via
\textcircled{\scriptsize{2}} 1-methylcyclohexa-1,3-diene to form
\textcircled{\scriptsize{3}} 1-methylcyclohexa-1,4-diene, with the manipulated hydrogen atom indicated in red.}
\label{fig:ReactionPathLewis}
\end{figure}

\begin{figure}[htb]
\includegraphics[trim=0 30 0 70, clip,scale=0.6]{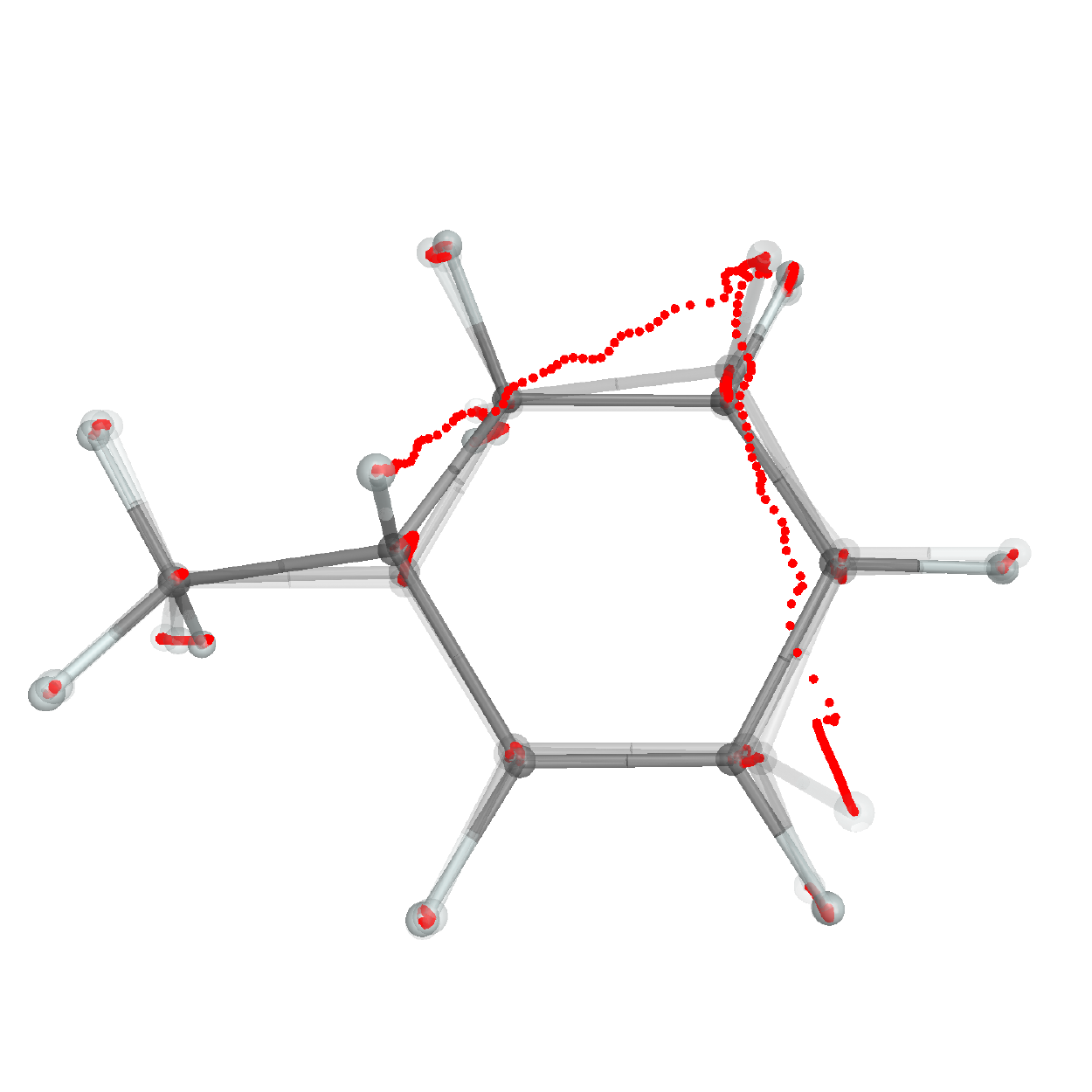}\\
\includegraphics[scale=1.2]{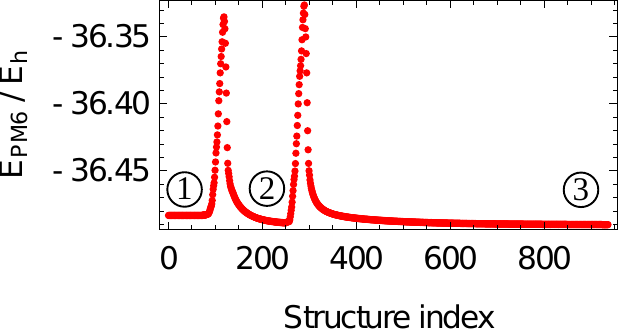}
\caption{
  Representation of $936$ data points (of dimension $3N\,{+}\,1=52$) obtained from $15.6$ seconds of interactive reactivity exploration of the chemical system presented in Fig.\,\ref{fig:ReactionPathLewis}.
  Top: Paths of the atoms along the sampled structure sequence.
  Each dot represents the position of one atom at each of the $936$ sampled structures.
  The chemical structures \textcircled{\scriptsize{1}}, \textcircled{\scriptsize{2}}, and \textcircled{\scriptsize{3}} of Fig.\ \ref{fig:ReactionPathLewis} are shown in decreasing opacity.
  Bottom: associated total electronic energies.
}
\label{fig:RawPath}
\end{figure}

We discuss the different processing steps of our algorithm at the example of (\textit{R})-5-methylcyclohexa-1,3-diene.
We explored two consecutive hydrogen shift reactions with the Parametrized Method 6 (PM6),\cite{stewart2007} for which we chose a spin-unrestricted framework with tailored SCF convergence acceleration,\cite{muehlbach2016a} forces generated with the mediator strategy,\cite{vaucher2016a} and orbital steering features.\cite{vaucher2017a}
The stable structures of this exploration path are depicted as Lewis structures in Fig.~\ref{fig:ReactionPathLewis} .
In the illustrations of this work, we will indicate the three stable structures by \textcircled{\scriptsize{1}}, \textcircled{\scriptsize{2}}, and \textcircled{\scriptsize{3}}, for the reaction sequence \textcircled{\scriptsize{1}} $\rightarrow$ \textcircled{\scriptsize{2}} $\rightarrow$ \textcircled{\scriptsize{3}}.
The chemical system consists of $N=17$ atoms, of which one hydrogen atom was moved around with a force-feedback haptic device with simultaneous structural relaxation to accomplish the two reaction steps. (For raw data of the recorded structure sequence, see the Supplementary Information.)
All atomic movements are shown in Fig.~\ref{fig:RawPath} together with the associated energies; there, the energy plot clearly shows the two distinct reaction steps.

We emphasize that our algorithm is general and can be applied to more complex examples and more accurate methods.

In Section~\ref{sec:bsplines}, we review B-spline curves and discuss their benefits for describing molecular paths.
Then, Section~\ref{sec:preconditioning} explains how a transformation is implemented to generate a continuous description of reaction paths with associated energies.
In Section~\ref{sec:processing}, we show how the reactivity information is extracted to generate a reaction network.

\section{Basis-Spline curves to describe molecular paths}\label{sec:bsplines}

In this section, we discuss B-spline curves and highlight their salient features for our path processing approach.
B-spline curves are a broad topic and we present the associated equations in their most general form.
For further details, we refer to the literature for B-spline curve evaluation, generation, manipulation, and analysis techniques (see Ref.~\citenum{piegl1997a} and references therein).

B-spline curves are univariate, parametric functions $\Curve(u)$ composed of $n$ piecewise polynomial segments of a specific degree $p$ with $1\leq p \leq n$.
The polynomial segments are attached so that the curve and its $p\,{-}\,1$ first derivatives are continuous.
The piecewise definition makes B-spline functions particularly flexible\cite{farin2002b} allowing for interpolation or approximation of complex-shaped and high-dimensional data, while maintaining a low polynomial degree.\cite{whitehorn2013a}
Also, polynomials represent a memory-efficient way of storing spatial information as few polynomial coefficients suffice to describe complex shapes.\cite{saux1998a,saux1999a}
Still, evaluations or spatial manipulations can be executed rapidly because only local polynomial segments must be considered.\cite{deboor1972a,boehm1984a}

\begin{figure}
\includegraphics[width=\linewidth]{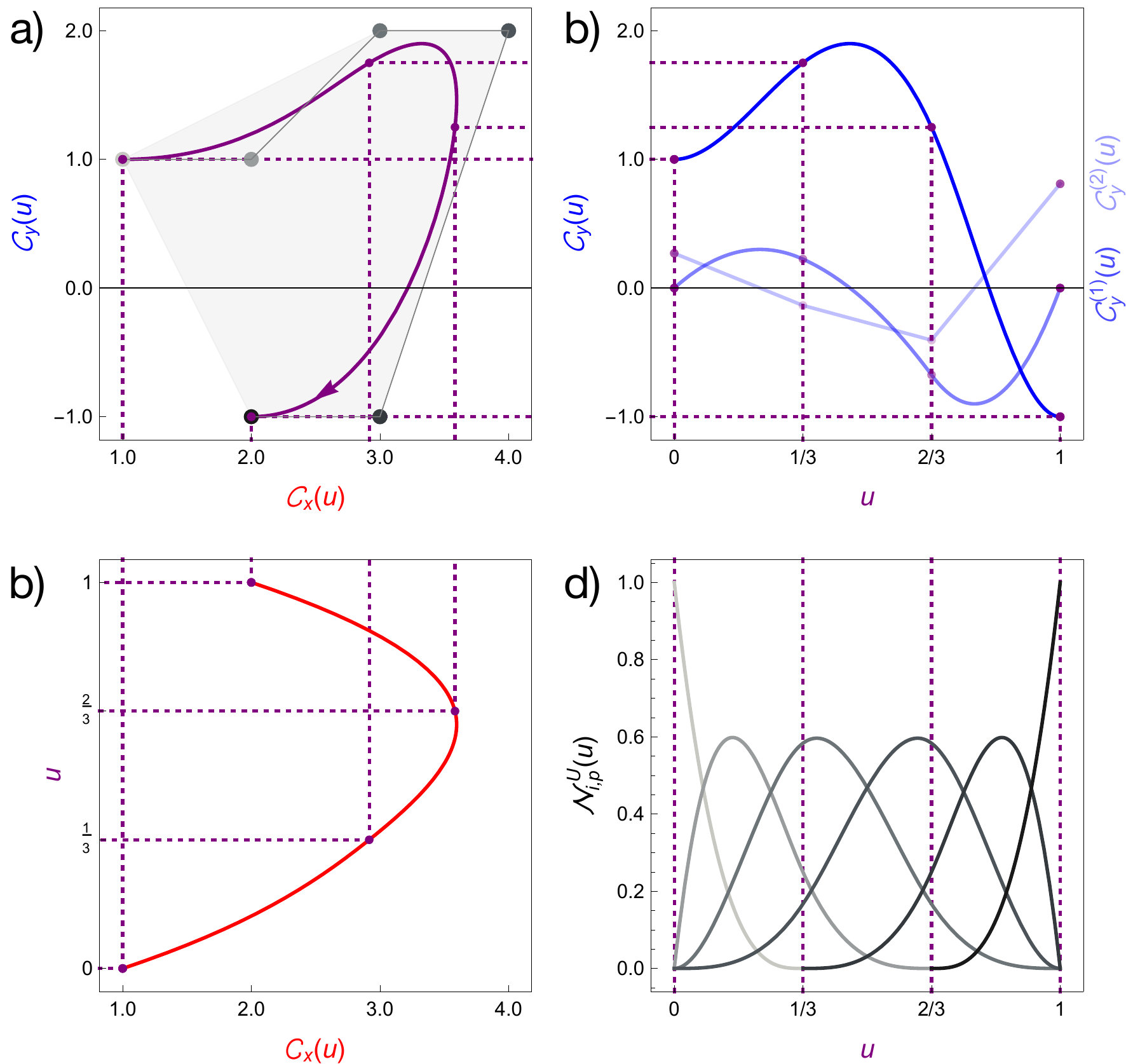}
\caption{
  Two-dimensional, cubic ($p\,{=}\,3$) B-spline curve $\Curve(u)$ parametrized on the knot vector
  $\Uvec\,{=}\,\{0,0,0,0,1/3,2/3,1,1,1,1\}$
  and geometrically defined by $(n\,{+}\,1\,{=}\,6)$ control points.
  \textbf{a})
  Parametric plot of $\CurveOneDim_y(u)$ against $\CurveOneDim_x(u)$ with
  the B-spline curve  $\Curve(u)= \{\CurveOneDim_x(u),\CurveOneDim_y(u)\}^\transp$ as a solid, purple line,
  the control points as gray-toned, filled circles connected by gray solid lines representing the control polygon;
  \textbf{b})
  plot of $\CurveOneDim_y(u)$ against $u$ as a solid, blue line with knots in purple and its scaled derivatives $\CurveOneDim^{(1)}_y(u) /10$ and $\CurveOneDim^{(2)}_y(u)/100$ with decreasing opacity;
  \textbf{c})
  plot of $u$ against $\CurveOneDim_x(u)$ as a solid, red line with knots in purple;
  \textbf{d})
  plot of the non-zero parts of the $(n+1=6)$ basis functions $\Nbas_{i,p}^{\Uvec}(u)$ as solid lines in different shades of gray.
}
\label{fig:BSpline2D}
\end{figure}

The B-spline curve function $\Curve(u) = \Curve^{(0)}(u)$,  which we define to be parametrized over the domain $u\in[0,1]$, and its derivatives $\Curve^{(k)}(u)$ of order $k$, can be written as a linear combination of B-splines $\Nbas_{i,p-k}^{\Uvec^{(k)}} (u)$,
\begin{align}
\Curve^{(k)}(u) &\equiv\frac{\partial^k \Curve^{(0)}(u)}{\partial u^k}
=\sum_{i=0}^{n-k}
\Nbas_{i,p-k}^{\Uvec^{(k)}} (u)\, \Pvec_i^{(k)},
\label{eq:BSplineCurveGeneral}
\end{align}
defined on a so-called knot vector $\Uvec^{(k)}$ and multiplied by control point vectors $\Pvec_i^{(k)}$.
The definitions of $\Nbas_{i,p-k}^{\Uvec^{(k)}} (u)$, $\Uvec^{(k)}$, and $\Pvec_i^{(k)}$ are given in Appendix~\ref{appendix:bsplines}.
An example of a two-dimensional B-spline curve is shown in Fig.~\ref{fig:BSpline2D}.

In this work, we describe sequences of structures containing $N$ atoms by $3N+1$-dimensional B-spline curves, where the additional dimension corresponds to the electronic energy.
The approximation as continuous, parametrized curves fits very well to the concept of a reaction coordinate in chemistry, which considers a continuous motion of atoms starting and ending in equilibrium structures.
Additionally, it allows for automatically interpolating in regions where few samples were recorded, and discarding duplicated data in densely sampled regions of the coordinate space.
The accessibility of the derivatives and the B-spline properties allow for an unconditionally convergent root-finding algorithm\cite{morken2007a} which we employ later in this work to analyze the curve in the energy dimension.
By finding roots $r_{z}$ and root intervals $[r_{a},r_{b}]$ in $\Curve^{(1)}(u)$, we can determine stationary points and stationary intervals in the original curve $\Curve^{(0)}(u)$.
Furthermore, B-spline curves can be split\cite{boehm1980a} and merged,\cite{tai2003a} which will be required in processing steps discussed later in this work.

\section{Preconditioning: Generation of continuous paths from sequences of structures}\label{sec:preconditioning}

To obtain continuous B-spline curves from a sequence of
$M$ sampled structures $\mathbf{R}\,{=}\,\{\Rvec_1,\cdots,\Rvec_M\}$ (with $\Rvec_f\,{=}\,\{x_{f1},y_{f1},z_{f1},\cdots,x_{fN},y_{fN},z_{fN}\}$ being $3N$-dimensional vectors constituted by the Cartesian atomic coordinates)
and associated energies $\boldsymbol{E}\,{=}\,\{E_1,E_2,\cdots,E_M\}$,
two preconditioning steps are required: polyline simplification, which eliminates redundant points, and penalized B-spline least-squares fitting, which generates a smooth approximating curve from the remaining points.
The preconditioning is important for the subsequent path-processing steps, because their computational cost is determined to a large extent by the quality of the fit.

In the case of interactive reactivity exploration, the structure sequences are recorded with a constant update frequency of 60~Hz.
The parts of the molecular system not directly manipulated by an operator evolve following the negative electronic energy gradient, $-\nabla_i E_\mathrm{el}(\Rvec)$ for each atomic nucleus $i$, according to a gradient descent minimization.
As a result, potential wells and low-energy regions tend to accumulate redundant points, especially when the operator is not conducting any manipulation.
Removing such points is beneficial for the least-squares fit of the second preconditioning step and decreases the amount of data that need to be stored.
By contrast, regions with steep gradients are more sparsely sampled.
This is often the case for the path regions around transition states.
In such sparsely sampled regions, points must be retained so that the representation of the path as a B-spline delivers a smooth structural change.
Note that these observations also hold for a configuration space sampling by moderate-temperature AIMD.

\subsection{Polyline simplification}
To remove redundant points from a sequence of structures, we consider the line segments between consecutive points and connect them to form a polyline. The polyline is then reduced by a simplification algorithm.
For this, we chose the Ramer--Doulgas--Peucker (RDP) algorithm\cite{ramer1972a,douglas1973a} in a modified formulation (as detailed below) specific to sequences of molecular structures.
The algorithm simplifies a polyline by removing points without changing the shape of the polyline within some given threshold.

The standard RDP algorithm starts with a single line segment $\Lcalvec$ connecting the first point $\Rvec_a$ and the last point $\Rvec_b$ of the original path.
The algorithm then searches the point $\Rvec_i$, $a<i<b$, in the original path with the maximal distance, $d^{\perp}(\Rvec_i, \Lcalvec)$, to the line $\Lcalvec$.
The calculation of the point-line distance $d^{\perp}(\Rvec_i, \Lcalvec)$ is detailed in Appendix~\ref{appendix:rdp}.
If $d^{\perp}(\Rvec_i, \Lcalvec)$ is above a threshold parameter $\varepsilon$, the original path will be split at $\Rvec_i$.
Then, the point $\Rvec_i$ is included in the simplified polyline and introduces two new line segments on which the algorithm can be reapplied.
This recursion results in a binary tree of line segments.
The recursion is stopped at segments where the maximal point-line distance is smaller than $\varepsilon$. In this case, all interior points of the original path are discarded, which simplifies the polyline.

For sequences of molecular structures, $\varepsilon$ must have the dimension of a length in $3N$ dimensions.
This suggests relating the choice of $\varepsilon$ to the number of atoms in some way.
In our case, the algorithm must recognize structures corresponding to a significant collective motion of all atoms, as well as a significant motion of one single atom, as important structures.

In order to cover both cases, we modified the RDP algorithm to consider two thresholds, $\varepsilon_\mathrm{c}$ and $\varepsilon_\mathrm{s}$.
We define $\varepsilon_\mathrm{s}$ as the single-atomic displacement threshold.
A structure will be kept in the simplified polyline if the distance for any single atom to the line $\Lcalvec$ exceeds $\varepsilon_\mathrm{s}$.
We define $\varepsilon_\mathrm{c}$ as the collective threshold;
a structure will also be kept if the root mean square distance of all atoms to the line $\Lcalvec$ exceeds $\varepsilon_\mathrm{c}$.

\begin{figure}[htb]
\includegraphics[trim=0 30 0 70, clip, scale=0.6]{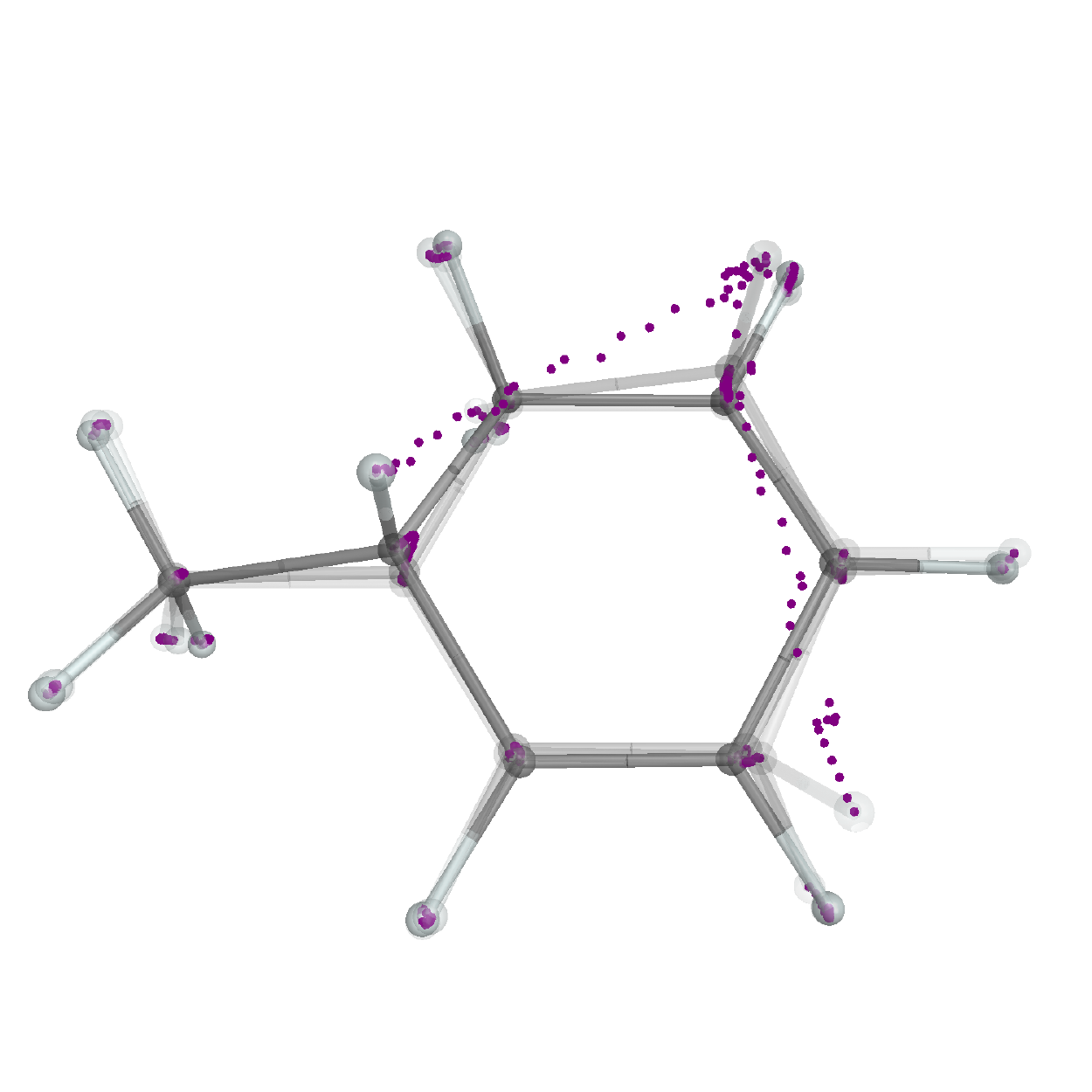}\\
\includegraphics[scale=1.2]{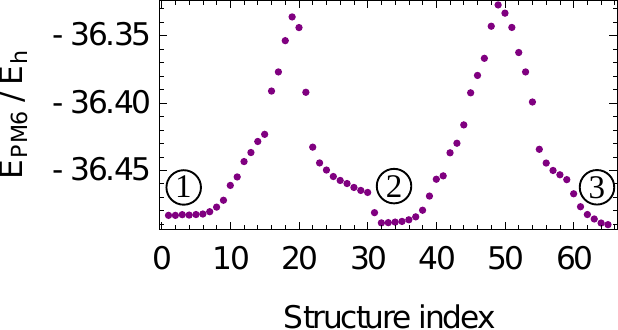}
\caption{
  65 data points obtained after RDP simplification of the raw path shown in Fig.~\ref{fig:RawPath}.
  Top: Paths of the atoms;
  each dot represents the position of one atom at each of the $65$ structures.
  The chemical structures \textcircled{\scriptsize{1}}, \textcircled{\scriptsize{2}}, and \textcircled{\scriptsize{3}} of Fig.\ \ref{fig:ReactionPathLewis} are shown in decreasing opacity.
  Bottom: associated total electronic energies.
}
\label{fig:SimplifiedPath}
\end{figure}

The result of applying the algorithm with $\varepsilon_\mathrm{s} = 10^{-2}\,\text{\AA}$ and $\varepsilon_\mathrm{c} = 5\times10^{-3}\,\text{\AA}$ is shown in Fig.~\ref{fig:SimplifiedPath}.
The raw exploration path consisting of 936 points is simplified to 65 points.
The reduced number of data points facilitates the further processing and storage of the exploration data.

\subsection{Penalized Basis-Spline Least-Squares Fitting}
\label{sec:PenalizedLeastSquares}

In the next pre-conditioning step, we approximate the discrete simplified exploration path by a cubic B-spline curve by penalized least-squares fitting (see Appendix~\ref{appendix:fitting} for details).
After polyline simplification, the points are distributed homogeneously over the interval in which the B-spline fit is performed.
This is crucial to achieve an appropriate curve representation because it avoids a bias in the fit that would be caused by uneven sampling.

The approximation of the exploration path by a cubic B-spline curve is of central importance for our path processing scheme as it allows us to include the total electronic energy as an additional dimension in the curve.
To include the energy, we construct a $(3N{+}1)$-dimensional exploration path vector
\begin{align}
\Qvec_f= \{E_f,x_{f1},y_{f1},z_{f1},\dots,x_{fN},y_{fN},z_{fN}\}^\text{T}
\end{align}
containing the energy as its first dimension.
Accordingly, the cubic B-spline fit to the exploration path $\Qmat\,{=}\,\{\Qvec_0,\Qvec_1,\dots,\Qvec_m\}$ then yields a $(3N{+}1)$-dimensional, $C^2$-continuous, parametric curve function $\boldsymbol{\mathcal{C}}(u)$.
Here, we benefit from the fact that the energies of the simplified exploration paths are already calculated so that no additional electronic structure calculations are required in the following steps.

A cubic B-spline curve including the energy holds the substantial advantage of obtaining continuous energy-derivatives $\mathcal{C}_0^{(k)}(u)$ up to $k=2$ along the entire path without conducting a single additional electronic structure calculation.
This enables us to perform a near-realtime curve analysis which is employed in the network construction algorithm to determine stationary points and intervals in the profile of the total electronic energy.
The penalized least-squares fitting further eliminates artificial oscillations or ones that may be introduced manually by the operator during exploration.

Further details about the penalized least-squares fitting algorithm applied here can be found in Appendix~\ref{appendix:fitting}.

\begin{figure}[htb]
\includegraphics[trim=0 30 0 70, clip, scale=0.6]{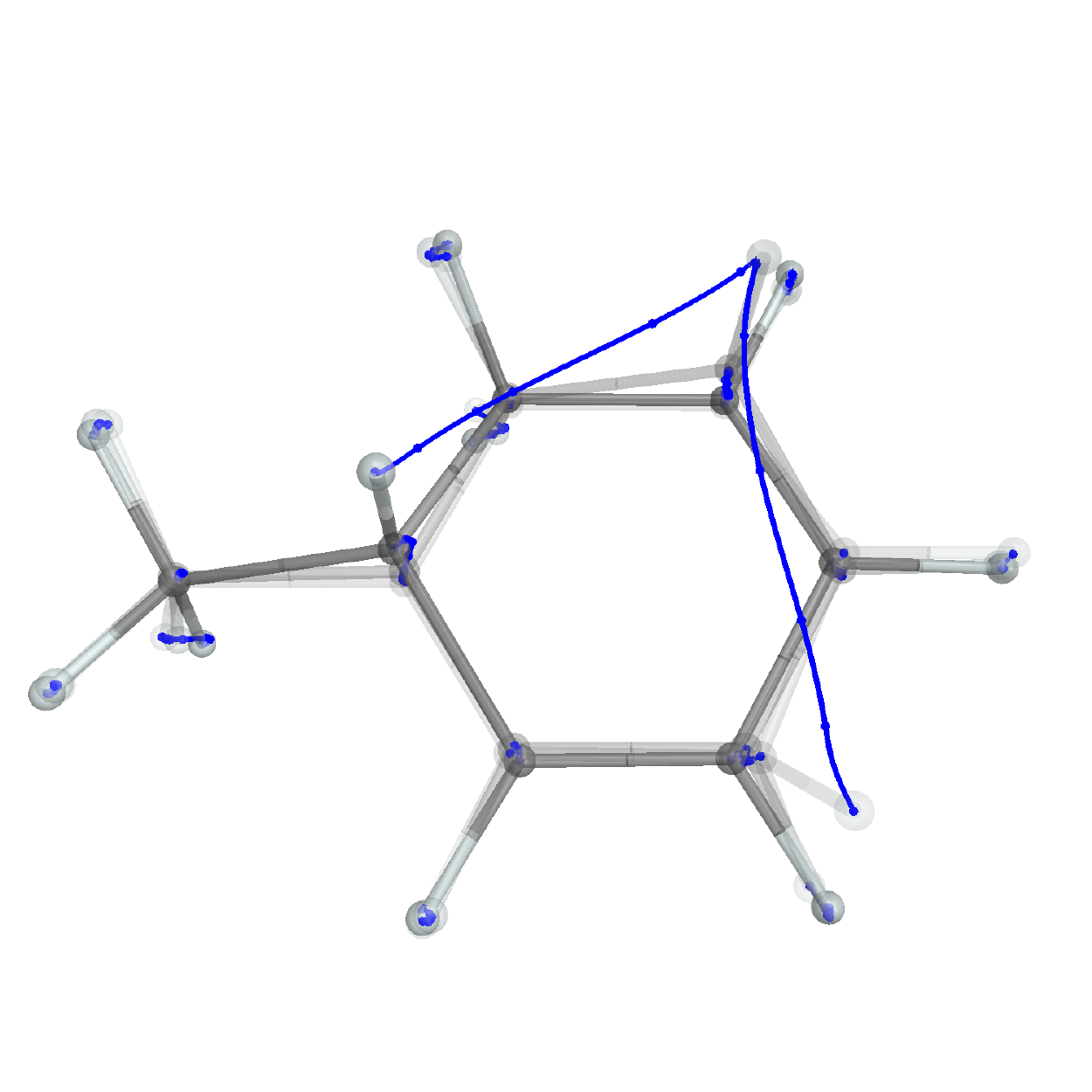}\\
\includegraphics[scale=1.2]{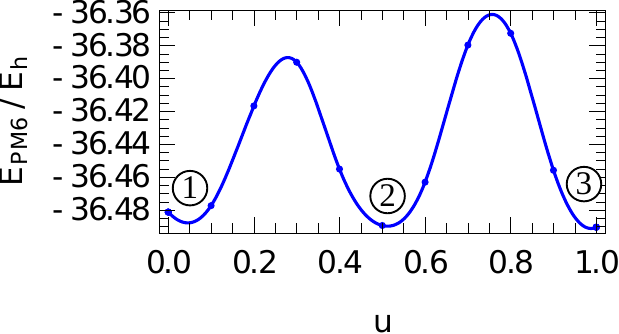}
\caption{
  Path approximated as a B-spline curve obtained from the simplified path shown in Fig.~\ref{fig:SimplifiedPath}.
  Top: Paths of the atoms;
  the chemical structures \textcircled{\scriptsize{1}}, \textcircled{\scriptsize{2}}, and \textcircled{\scriptsize{3}} of Fig.\ \ref{fig:ReactionPathLewis} are shown in decreasing opacity.
  Bottom: associated total electronic energies plotted against the parameter $u$.
}
\label{fig:SimpFitted}
\end{figure}

Converting the sequence of structures from our example delivers $13$ control points connected by B-spline curves containing the corresponding energies, as shown in Fig.~\ref{fig:SimpFitted}.
Additionally, the data reduction further reduces the memory requirements to store the molecular paths.

\section{Path Processing}\label{sec:processing}

In this section, we introduce a path-processing scheme for the construction of a reaction network from a fixed-size, preconditioned exploration path.
The scheme presented here processes exploration paths described by B-spline curves and extracts its stable molecular structures and elementary reaction steps to form a reaction network. 
This scheme can be extended for on-the-fly processing by buffering, which we discuss later.

At first, an exploration network is constructed from the fixed-sized, preconditioned exploration path.
It is similar to the target reaction network, but consists of unrelaxed nodes and edges representing candidates for stable chemical structures and the elementary reaction steps interconnecting them, respectively.
Besides candidates for elementary reactions, the exploration path can also contain path segments representing pure translations or rotations of non-interacting, dissociated fragments.
To distinguish such segments from reactions, they will be represented by a special edge type in the exploration network.
The exploration network must then be relaxed to yield the unique nodes and edges of the target reaction network.

\begin{figure*}[tb]
\centering
\includegraphics[width = 0.7\linewidth]{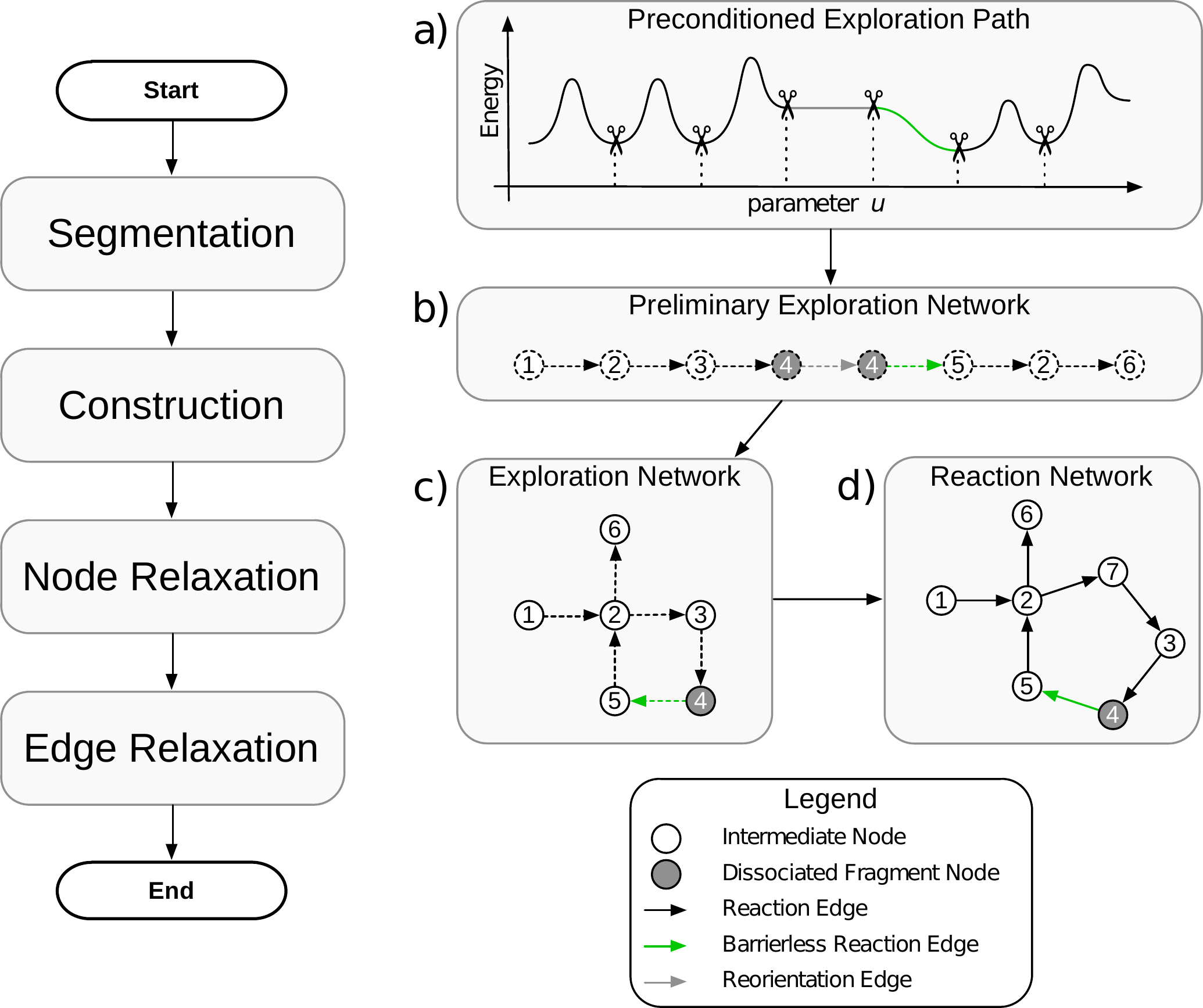}
\caption{
  Schematic representation of the path-processing scheme with
  a) segmentation of the preconditioned exploration path,
  b) construction of the exploration network as unrelaxed nodes and directed edges indicated by dashed circles and arrows, respectively,
  c) cross-linking due to intermediate \textcircled{\scriptsize{2}} and clustering of dissociation node \textcircled{\scriptsize{4}} in the exploration network after relaxation (relaxed nodes as solid circles) and comparison, and
  d) reaction network after edge relaxation showing the unconsidered intermediate \textcircled{\scriptsize{7}}.}
\label{fig:PathProcessingScheme}
\end{figure*}

Accordingly, several processing steps are necessary to obtain the optimized reaction network.
Each one will be discussed in the following.
An schematic overview of these steps is presented in Fig.~\ref{fig:PathProcessingScheme}.a-d.

\subsection{Splitting the Explored Path at Stable Structures}\label{subsec:segmentation}

Before constructing the exploration network, we must identify the candidates for stable chemical structures and the interconnecting reaction events based on the exploration energy.
Once found, the candidates for stable chemical structures define where to cut the preconditioned exploration path into segments.
Accordingly, each segment then represents two candidate nodes connected by a candidate edge in the exploration network graph.

Generally, a system will be defined as stable if it returns to its initial equilibrium state after a small pertubation in one or more of its variables.
For a chemical structure, this condition is strictly fulfilled only for local minima on the PES.
However, it is reasonable for our purposes to also consider points in very flat regions of the PES, indicated by an interval of stationary exploration energy, as stable.

After the preconditioning step, redundant structures have been eliminated and it is thereby ensured that stationary energy intervals cannot be caused by accumulated, identical structure vectors.
Accordingly, stationary energy intervals describe regions of no interaction occurring when molecular fragments are translated or rotated independently at sufficiently large distances, which corresponds to separated or dissociated chemical species (for instance, two distinct molecules).
Here, we will consider species to be dissociated when translation or rotation of the fragments can be viewed as a movement in a sub-region of the PES that is flat with respect to a given threshold.
With the ability to detect independent species within a molecular structure, we can identify association, dissociation, or substitution reactions in the reaction network.

The start and end points of stationary intervals mark the re-entry points in the interaction region.
Together with minima, they can be conveniently detected by B-spline curve analysis\cite{morken2007a} of the exploration energy dimension and are employed for cutting the preconditioned exploration path into segments.
This is shown in Fig.~\ref{fig:PathProcessingScheme}.a.

\subsection{Exploration Network Construction}

In the next step we take the path segments obtained from splitting the exploration path for the construction of the exploration network.
Such a segment is represented in the network graph as two nodes connected by an edge.
Accordingly, the segmented exploration path translates into a linear sequence of nodes and edges as depicted in Fig.~\ref{fig:PathProcessingScheme}.b constituting the initial state of the exploration network.

This preliminary exploration network is composed of different node and edge types, which we define in the following.
As mentioned already in the previous section, a node represents either a stable chemical system or a candidate for it.
We introduce two different node types to differentiate between truly stable intermediates and dissociated fragments representing minima and points in flat regions of the PES, respectively.
Accordingly, we denote them as intermediate nodes (I-nodes) and dissociated-fragment nodes (DF-nodes).

An edge always connects two nodes and represents a possible path converting the two stable systems at its ends into each other.
Such a segment constitutes an initial guesses for a MEP that connects two minima on the PES.
Again, we must distinguish different types of path segments.
If the sequence results in a change of the total electronic energy and molecular structure, it will be depicted as a reaction edge (R-edge).
If the energy remains constant along the path segment, we will call it a reorientation and represent it as a reorientation edge (RO-edge) in the network.
Reorientations occur when dissociated, non-interacting fragments are translated or rotated, while being far away from each other so that the interaction energy between them is negligibly small.
Also, we must differentiate between reactions that include an activation barrier and well-defined transition state, and reactions with monotonically increasing or monotically decreasing energies (such as barrierless association reactions).
Accordingly, we introduce a further edge type and call it barrierless reaction edge (BLR-edge).

Altogether, these definitions allow us to reflect chemical reactivity in our network.
Reactions can transform I-nodes into other I-nodes, which corresponds to a reaction of the general type \mbox{A $\xrightleftharpoons{}$ B}.
I-nodes can also dissociate into non-interacting fragments represented by a DF-node which corresponds to \mbox{A $\xrightleftharpoons{}$ B + C}.
The reverse reaction is then an association of non-interacting fragments within a DF-node into a I-node.
Finally, substitution reactions have DF-nodes both at the start and end of the reaction which reads as \mbox{A + B $\xrightleftharpoons{}$ C + D}.

With the exception of the transformation of an I-node into another I-node, dissociation and substitution reactions can be barrierless as well.
Reorientations can only occur between dissociated fragments and therefore must be connected by two DF-nodes.
They involve no change in the chemical system and lead to chemically identical, non-interacting fragments that differ only by their relative position or orientation in space.

As mentioned earlier, nodes and edges of the exploration network are only initial guesses, or candidates, for the optimized network nodes and edges.
In the following, we indicate such nodes and edges by prepending the letter 'g' to the corresponding names.
Their relaxation is discussed in the next two sections.

\subsection{Node Relaxation}\label{subsec:node-relaxation}

To obtain the nodes of the target reaction network, the candidate intermediate nodes (gI-nodes) and candidate dissociated fragment nodes (gDF-nodes) must be relaxed.
This is achieved by geometry optimization of the corresponding molecular structures.

Each node relaxation generates an additional structure sequence representing the relaxation path from the initial guess to a true stable point on the PES.
To maintain the continuous connection between a relaxed node and the remaining network, the relaxation path is preconditioned and merged to every edge connected to the original node by B-spline curve merging.\cite{tai2003a}

It is crucial for the modeling of a reaction network that each individual, stable chemical system is represented by a single, unique node.
To assure the uniqueness of nodes in the reaction network, each relaxed node must be compared with all other relaxed nodes in the network to identify equivalent chemical structures.
This can be achieved, for instance, with the help of metrics based on the root mean square deviation (RMSD).
Thereby, invariance with respect to rotation and translation must be respected, and in case of dissociated fragments each one must be considered separately.
More complex algorithms taking into account the permutations of identical atoms can also be applied.\cite{sadeghi2013a,de2016a}

If two nodes are found to be chemically identical, both nodes must be merged into a single node.
If the two nodes are not adjacent, the unification will connect different parts of the network with each other, which we denote as cross-linking.
In Fig.~\ref{fig:PathProcessingScheme}.b, this is the case for the nodes  labeled as \textcircled{\scriptsize{2}}.
If two nodes are already connected, we denote the unification as clustering.
In Fig.~\ref{fig:PathProcessingScheme}.b, this is the case for the nodes labeled as \textcircled{\scriptsize{4}}.
The resulting exploration network after node relaxation, comparison, cross-linking and clustering is shown in Fig.~\ref{fig:PathProcessingScheme}.c.

\subsection{Edge Relaxation}

After the nodes have been relaxed, each edge connects two stable points on the PES.
However, the edge itself is still a guess and is likely to deviate from the MEP and may even include other stable points.
Therefore, every candidate edge must be relaxed to verify the direct connection between its source and destination node.
Furthermore, relaxation allows us to find a possible transition state (R-edge), ensure its absence (BLR-edge), or demonstrate the flatness of the PES region (RO-edge).

If the edge relaxation reveals one or more intermediates, the edge is split accordingly and candidate nodes are added for the intermediates.
The new nodes and edges must then be refined as well.

In other cases, an unrelaxed gR-edge can turn out to be a BLR- or RO-edge after relaxation and must be replaced in the target-network accordingly.
For the latter, the nodes at the start and end must be replaced by DF-nodes.
Similarly to gR-edges, also gBLR- and gRO-edges must be confirmed by relaxation.
If multiple, subsequent RO-edges occur in a network, they can be substituted by a single one.

It can also occur that two relaxed nodes are connected by multiple edges.
In this case, comparison of the paths allows to remove duplicates; otherwise, they correspond to different local MEPs and are retained.

To obtain MEPs and transition state estimates, we apply ReaDuct, a new MEP optimization method for transition paths described by parametrized curves.\cite{vaucher2018a}
Note that other methods, such as the nudged elastic band (NEB) method\cite{jonsson1998a,henkelman2000a,henkelman2000b} or the string method\cite{weinan2002a,weinan2005a} can also be applied for edge relaxation.

\begin{figure*}
\includegraphics[trim=0 30 0 70, clip, scale=0.6]{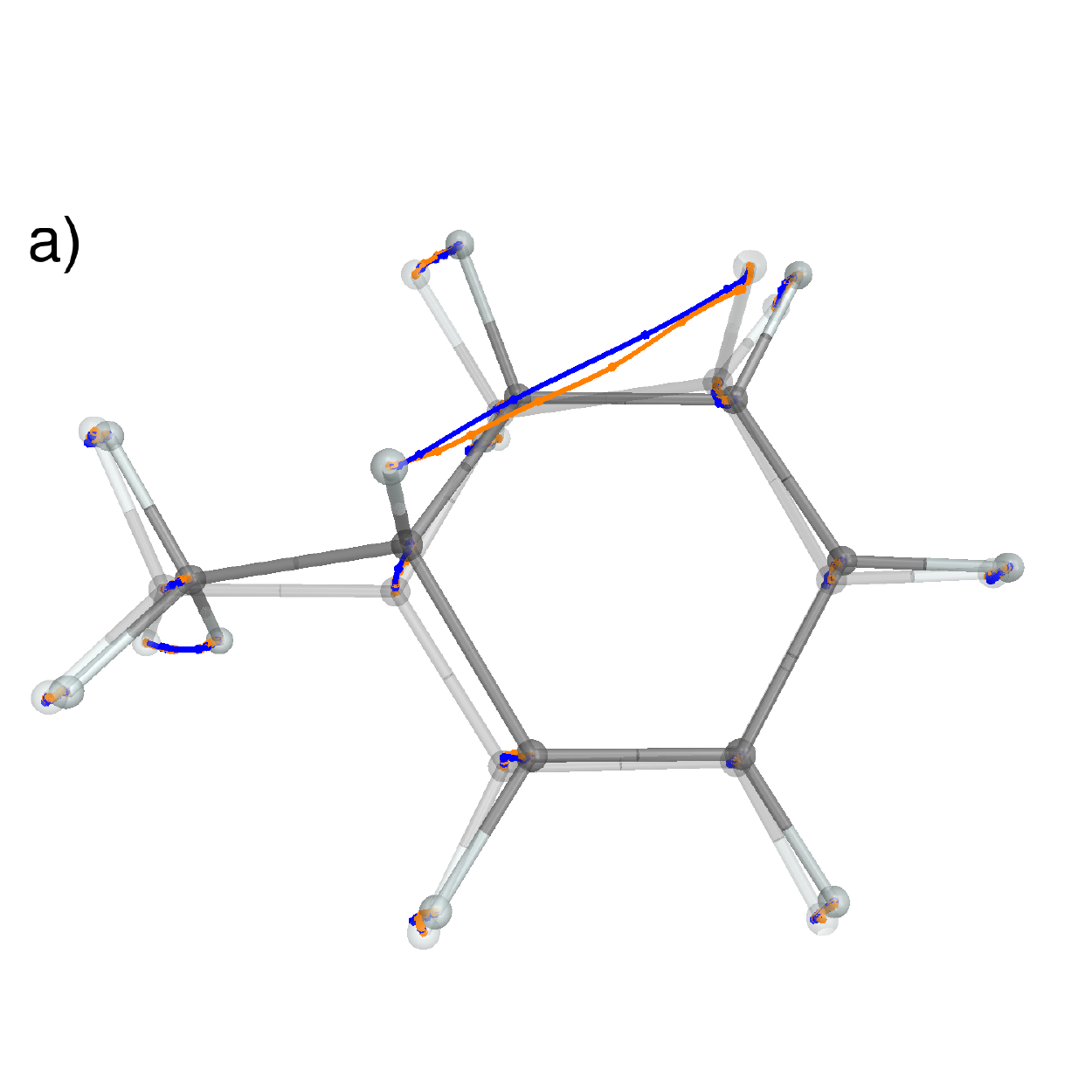}\hspace{1.5cm}
\includegraphics[trim=0 30 0 70, clip, scale=0.6]{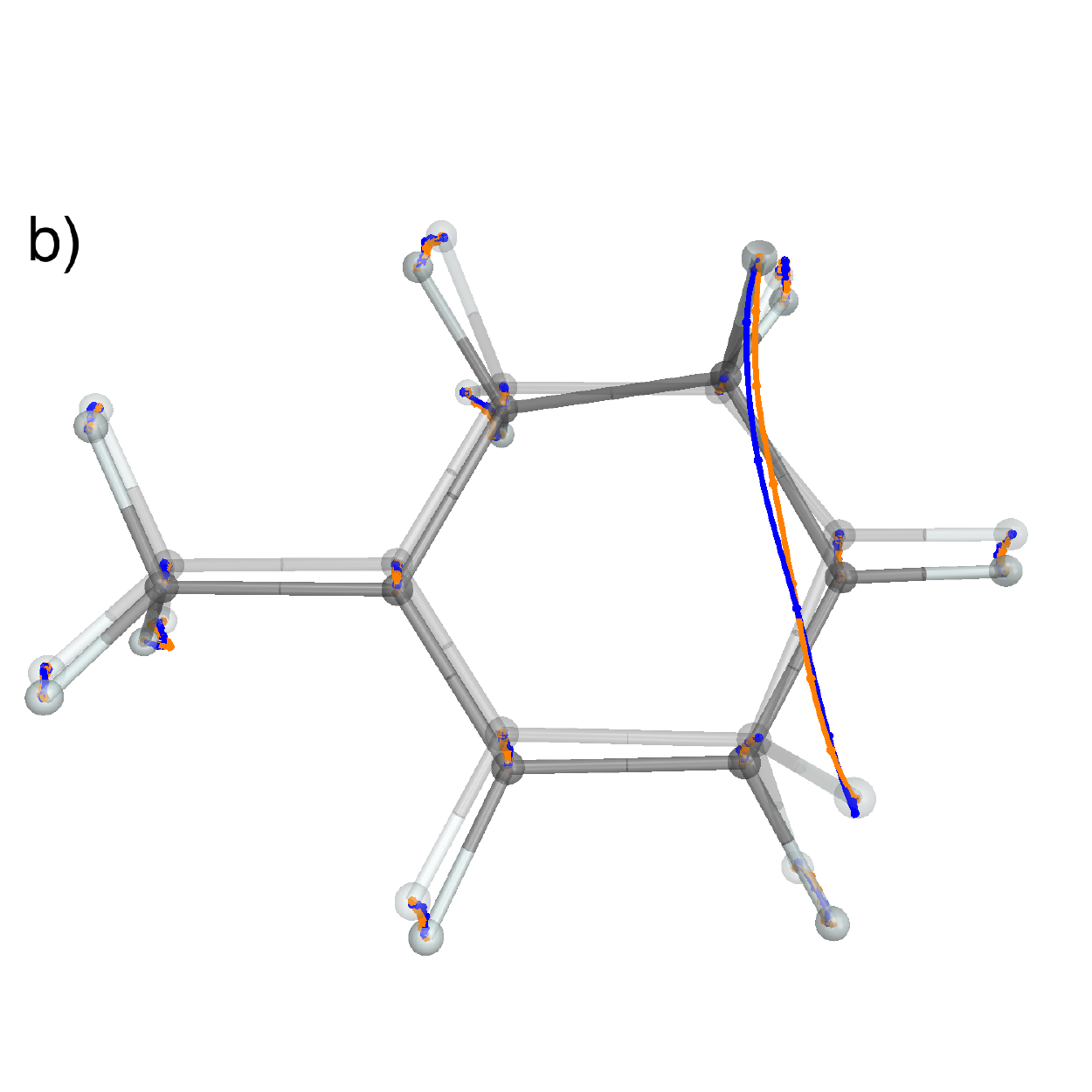}
\\
\includegraphics[scale=1.2]{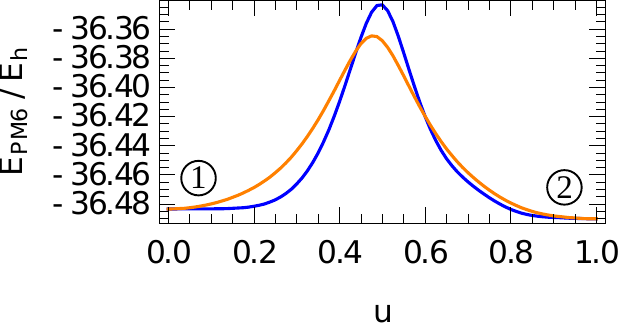}\hspace{1.5cm}
\includegraphics[scale=1.2]{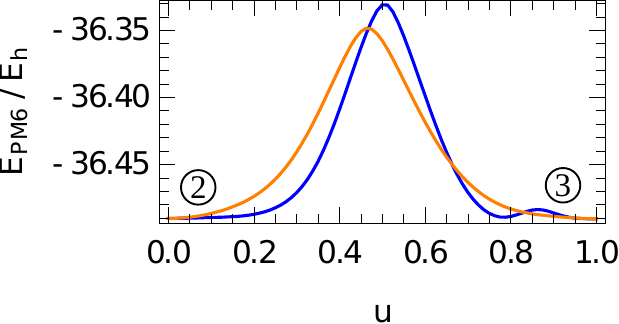}
\caption{
  B-spline curve segments of the a) first and b) second explored reaction step showing the exploration form \textcircled{\scriptsize{1}} to \textcircled{\scriptsize{2}} and \textcircled{\scriptsize{2}} to \textcircled{\scriptsize{3}}, respectively.
  The segments obtained by splitting of the preconditioned exploration path from Fig.~\ref{fig:SimpFitted} are indicated in blue, whereas the the ReaDuct optimization result is indicated in orange.
  Top: Paths of the atoms;
  Bottom: associated total electronic energies plotted against the parameter $u$.
}
\label{fig:EdgeRelaxation}
\end{figure*}

The ReaDuct method optimizes reaction paths by optimizing the parameters of continuous curves (in this work, B-spline curves).
This leads to a more natural formulation with less parameters, and is therefore more adapted to the automated optimization of reaction paths, which is essential for this work.
A detailed discussion of ReaDuct is beyond the scope of this work and will be presented elsewhere.\cite{vaucher2018a}
The application of the ReaDuct algorithm to our example changes the path followed by the atomic nuclei during the reactions and leads to lower activation energies, as can be seen in Fig.~\ref{fig:EdgeRelaxation}.

\section{On-the-fly path processing in a real-time Application}\label{sec:on-the-fly}

The protocol presented in Section~\ref{sec:processing} assumes that the complete path is known when its processing starts.
However, it can easily be adapted to process the exploration path on-the-fly.
Doing so bears the advantage to distribute the computational cost over the whole exploration duration and avoids a potentially expensive computation at the end of the exploration.
Furthermore, in an interactive setting, the program operator can, shortly after exploring some reaction, access the relaxed structures and elementary reactions and conduct further explorations accordingly.

To implement such an on-the-fly path processing, the continuous stream of structures and energies is divided into sets of some given size by buffering.
When the buffer is full, the corresponding set is preconditioned and merged, if applicable, to some remaining unprocessed B-spline from the previous buffer set.
Then, the resulting B-spline is processed as described earlier, but for the segment following the last discovered node.
This segment will be merged when the next buffer set must be processed.

\section{Conclusion and Outlook}

In this work, we detailed the conversion of a sequence of structures into a reaction network in the form of nodes (minimum structures) and edges (elementary reactions).
The sequence of structures could be generated, for instance, during an interactive reactivity exploration or a Born--Oppenheimer molecular dynamics simulation.

To achieve this, we first converted the explored path to a continuous curve represented by a B-splines.
This was achieved by a polyline simplification followed by penalized least-square fitting of the sequence of structures generated during the reactivity exploration.
This procedure produced a preliminary exploration path as a B-spline curve of dimension $3N+1$ that contains the coordinates of the $N$ atoms as well as one dimension for the energy.

Representing the exploration path (and the edges of the reaction network) as B-spline curves not only fits the concept of a continuous reaction coordinate, but it also makes splitting, merging, or manipulating path segments easier by applying well-established algorithms.

From the preliminary exploration path, candidates for the nodes and edges of the reaction network can be generated without additional single-point calculations.
The minimum structures and elementary reactions corresponding to the exploration are then obtained by relaxation of the candidate nodes and edges to form a reaction network.
In the reaction network, we introduced different types of nodes and edges to reflect the fact that some edges represent reorientation of molecules or barrierless reactions.
This allows for rapidly identifying different types of chemical reactions (association, dissociation, substitution, transformation).

To illustrate the steps of our algorithm, we created a reaction network for two consecutive hydrogen shift reactions in (\textit{R})-5-methylcyclohexa-1,3-diene.
In practice, the algorithms for B-spline curve manipulation can be executed efficiently and scale linearly with the number of dimensions.
Therefore, the size or complexity of chemical systems that can be studied with our algorithm can be much larger and is not limited to simple cases such as this example.

The full procedure is automated and all steps can be parallelized and executed without user input.
This also allows one to refine the exploration path on-the-fly and build up a reaction network interactively.
Additionally, our path-processing approach can be applied to refine a reaction network with more accurate quantum chemical methods.
It is also straightforward to combine reaction networks generated from sequences of structures with reaction networks obtained by automated approaches.\cite{haag2014a,simm2017a}

\section*{Acknowledgments}

This work was generously supported by ETH Research Grant ETH-20 15-1 and ETH Pioneer Fellowship Grant PIO-11-14-2.

\section{Appendix}

\subsection{B-splines}\label{appendix:bsplines} 
B-splines $\Nbas_{i,p-k}^{\Uvec^{(k)}} (u)$ are univariate, parametric, polynomial functions of degree $p$$-$$k$ which are non-zero only on certain sub-intervals of the domain (see Fig.~\ref{fig:BSpline2D}.d).
The sum over all B-spline functions for any parameter value $u\in[0,1]$ in the domain is equal to one and therefore they constitute a partition of unity.
They can be obtained from the Cox-de\,Boor-Mansfield recurrence relation\cite{cox1972a,deboor1972a} by
\begin{align}
\Nbas_{i,p-k}^{\Uvec^{(k)}}(u) =	& \,\, \eta_{i,p-k-1}^{(k)}(u)\, \Nbas_{i,p-k-1}^{\Uvec^{(k)}}(u) \nonumber\\
						&+ \left[1-\eta_{i+1,p-k-1}^{(k)}(u)\right]\, \Nbas_{i+1,p-k-1}^{\Uvec^{(k)}}(u)
\label{eq:CoxDeBoorMansfieldRecurrence}
\end{align}
with the zeroth-order B-splines defined as
\begin{align}
\Nbas_{i,0}^{\Uvec^{(k)}}(u) &= 	\begin{cases}
						1, & \text{if } u\in\left[u_i^{(k)},u_{i+1}^{(k)}\right)\,\\[1mm]
						0, & \text{otherwise}
						\end{cases}
\end{align}
and the prefactors
\begin{align}
\eta_{g,h}^{(k)}(u)&=	\begin{cases}
				0, 									& \text{if } u_{g+h+1}^{(k)} = u_{g}^{(k)}\\[1mm]
				\displaystyle \frac{u-u_g^{(k)}}{u_{g+h+1}^{(k)}-u_g^{(k)}},	& \text{otherwise}
				\end{cases}
\label{eq:BSplineBasisFactor}
,
\end{align}
where the values $u_i^{(k)}$ are called knots and are stored in the so-called knot vector $\Uvec^{(k)}$.

\subsubsection{Knot Vector} 

The knot vector $\Uvec^{(k)}$ contains the knots in ascending order
\begin{align}
\Uvec^{(k)} &=\left\{u_0^{(k)},\dots, u_{n+p+1-2k}^{(k)}\right\}, \quad u_{j}^{(k)}\leq u_{j+1}^{(k)},
\end{align}
and defines the parametrization of the curve.
The knots lie, like the parameter $u$, in the domain $[0,1]$ and indicate the points where the polynomial segments are attached to each other (see Fig.~\ref{fig:BSpline2D}.d).
Here, we only consider B-spline curves with knot vectors where the first and last $p\,{+}\,1$ knots contain the values $0$ and $1$, respectively ($u_0\,{=}\,\dots \,{=}\,u_{p}\,{=}\,0$ and $u_{n+1}\,{=}\,\dots \,{=}\,u_{n+p+1}\,{=}\,1$).
Such B-spline curves are called \emph{clamped}.
The remaining knots are spaced equidistantly over the domain $[0,1]$ so that
\begin{align}
u_j^{(k)} &= 	\begin{cases}
			0, & j\in\{0,\dots,p{-}k\} \\
			\displaystyle \frac{j-p+k}{n-p+1}, & j\in\{p+1-k,\dots,n{-}k\} \\
			1, & j\in\{n{+}1{-}k,\dots,n{+}p{+}1{-}2k\}
			\end{cases}
\label{eq:KnotVector}
\end{align}

\subsubsection{Control Point Vectors} 

The control point vectors $\Pvec_i^{(k)}$ can be of arbitrary dimensionality and define the B-spline curve spatially.
$\Pvec_i^{(0)}$  can be obtained from data points by curve approximation via penalized least-squares fitting\cite{eilers1996a} (as is done in Sec.\,\ref{sec:PenalizedLeastSquares}, see Appendix~\ref{appendix:fitting}) or interpolation.\cite{piegl1997a}

The control points of the $k$-th derivative can then be obtained by
\begin{align}
\Pvec_i^{(k)}&=	\begin{cases}
			\Pvec_{i}^{(0)}, & k=0 \\[2mm]
			\displaystyle \frac{p-k+1}{u_{i+p+1}^{(0)} - u_{i+k}^{(0)}} \left( \Pvec_{i+1}^{(k-1)} - \Pvec_{i}^{(k-1)} \right), & k>0\\[2mm]
			\mathbf{0}, & \hspace{-12mm}u_{i+p+1}^{(0)} = u_{i+k}^{(0)}
			\end{cases}
\label{eq:ControlPointsGeneral}
\end{align}

\subsection{De\,Boor Algorithm}

Faster evaluation of Eq.\,\eqref{eq:BSplineCurveGeneral} is possible by exploiting the de\,Boor Algorithm\cite{deboor1972a}
where the  B-spline curve equation reads
\begin{align}
\Curve^{(k)}(u) &= \boldsymbol{A}_{l,p-k}^{(k)}, \quad
u\in [u_l^{(k)},u_{l+1}^{(k)}).
\end{align}
$\boldsymbol{A}_{i,j}^{(k)}$ is given by
\begin{align}
\boldsymbol{A}_{i,j}^{(k)} &=
\begin{cases}
	\Pvec_i^{(k)}, & j=0 \\[2mm]
	\left(1-\alpha_{i,j}^{(k)}\right)\, \boldsymbol{A}_{i-1,j-1}^{(k)} + \alpha_{i,j}^{(k)}\,\boldsymbol{A}_{i,j-1}^{(k)}, & j>0
\end{cases}
\end{align}
with $i \in\{l-(p-j),\dots,l\}$ and
\begin{align}
\alpha_{i,j}^{(k)}&=\frac{u-u_i^{(k)}}{u_{i+(p-j)+1}^{(k)} - u_{i}^{(k)}}
.
\end{align}

The algorithm is computationally more efficient because fewer calculations are required to evaluate the B-spline curve and its derivatives for a given parameter $u$.
Even faster evaluation is possible by the algorithms presented in Ref.~\citenum{boehm1984a} but their advantage is only significant for polynomial degrees larger than three, $p>3$.

\subsection{Calculation of the point-line distance in the Ramer--Douglas--Peucker algorithm}\label{appendix:rdp}

Let
\begin{align}
\Lcalvec(t) =  \Ocalvec + t\cdot   \Dcalvec
\end{align}
be a parametrized, infinite line in $3N$-dimensional space defined by an origin vector $\Ocalvec$ and a direction vector $\Dcalvec$.
Accordingly, a line passing through the points $\Rvec_a$ and $\Rvec_b$ can be defined by $\Ocalvec=\Rvec_a$ and $\Dcalvec=\Rvec_b-\Rvec_a$.

The point-line distance $d^{\perp}(\Rvec_i,  \Lcalvec)$ between a point $\Rvec_i$ and the line $ \Lcalvec$ is then obtained by
\begin{align}
d^{\perp}(\Rvec_i,  \Lcalvec) = \lVert ( \Ocalvec- \Rvec_i) - ( \Ocalvec- \Rvec_i)\, \Dcalvec \cdot  \Dcalvec \rVert
\label{eq:PointToLineDistance}
,
\end{align}
with $( \Ocalvec\,{-}\,\Rvec_i)\, \Dcalvec \cdot  \Dcalvec$ being the projection of $ \Ocalvec\,{-}\,\Rvec_i$ onto the line and $\lVert\, \cdot\, \rVert$ denoting the Euclidean distance.
The Euclidean distance is determined in a $3N$-dimensional space by
\begin{align}
d(d_1,\dots,d_{3N}) &=  \left(\sum_{i=1}^{3N} d_i^2\right)^{1/2}
\label{eq:EuclideanDistance}
,
\end{align}
with $d_i$, $i\in\{1,\dots,3N\}$, being the differences in a single coordinate.

\subsection{Penalized Least-squares Fitting}\label{appendix:fitting}

The penalized least-squares fitting procedure\cite{eilers1996a,eilers2010a} determines a B-spline curve by approximating a series of given data points.
To obtain the approximating curve, one determines the control points $\Pvec^\mathrm{LS}_i$ minimizing the squared differences of the curve at equidistant parameters $\bar{u}_g$, $\Curve(\bar{u}_g)$, to the data points $\Qvec_g$.
The number of control points is not determined by the fitting procedure and must be chosen beforehand.
The objective function subject to minimization then reads
\begin{align}
\Pvec^\mathrm{LS} = \operatornamewithlimits{arg\ min}_{\Pvec} \left\{ \sum_{g=0}^{m} \left\lVert \Qvec_g - \sum_{i=0}^{n} \mathcal{N}_{i,p}^{\Uvec}(\bar{u}_g)\,\Pvec_i \right\rVert^2 \nonumber \right. \\
                    \qquad \qquad \left. + \lambda \sum_{i=\kappa}^{n} (\Delta^\kappa \Pvec_i)^2 \right\}
\label{eq:PenalizedBSplineCurveApproximation}
,
\end{align}
with the B-splines evaluated at the equidistant parameters
\begin{align}
\bar{u}_g=	\begin{cases}
		0, & g=0 \\
		g/n, & g\in\{1,\dots,m-1\} \\
		1, & g=m
		\end{cases}
,
\end{align}
the smoothing parameter $\lambda$, and the $\kappa$-th order, difference operator $\Delta^{\kappa} = \Delta(\Delta^{\kappa-1})$, $\Delta \Pvec_i = \Pvec_i - \Pvec_{i-1}$. \cite{eilers1996a,eilers2010a}
The objective function results in a system of $m+1$ linearly independent equations containing the $n+1$ control points $\Pvec_i$ as variables (see below for the matrix representation).

For a smoothing parameter $\lambda > 0$, a penalty term containing the $\kappa$-th order, finite
differences of the control points $\Pvec_i$ is included in Eq.\,\eqref{eq:PenalizedBSplineCurveApproximation}.
For equidistant parameters $\bar{u}_g$, the penalty is obtained from the difference operator $\Delta^{\kappa} = \Delta(\Delta^{\kappa-1})$ with $\Delta \Pvec_i = \Pvec_i - \Pvec_{i-1}$. \cite{eilers1996a,eilers2010a}
For $\lambda = 0$, the unpenalized least-squares fitting procedure is obtained\cite{eilers2010a}.

The penalty is important to prevent undesired oscillations in the approximating curve that occur if sharp turns or oscillations are present in the data points.
For application in the approximation of molecular paths, these oscillations would lead to distorted coordinates and oscillations in the energy resulting in wrong guesses for stable intermediates.
By penalizing large variations in adjacent control points, this oscillatory curve behavior can be successfully suppressed and leads to smooth approximating curves.

\begin{figure}
\includegraphics[width = \linewidth]{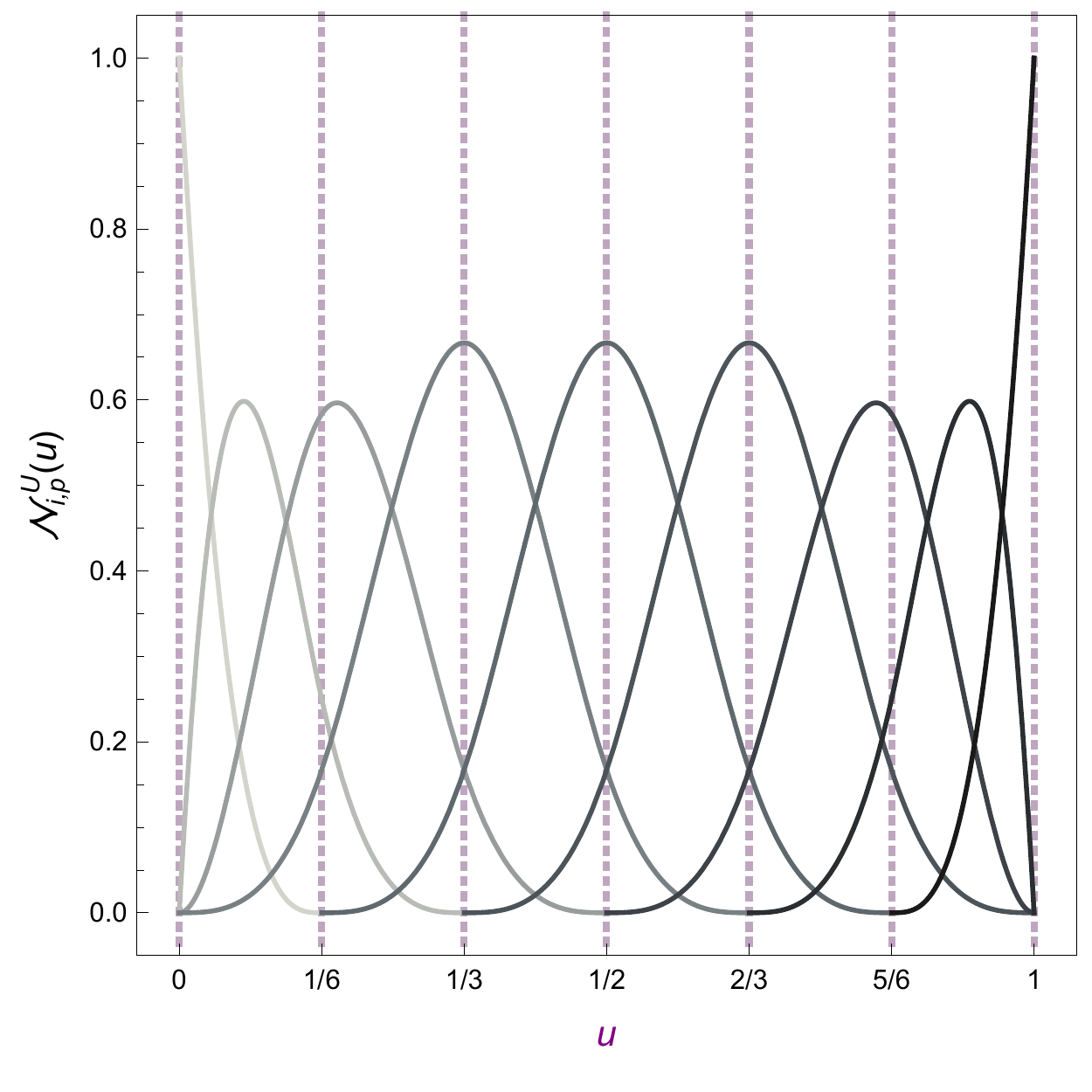}
\caption{B-splines for a cubic curve with $n=8$ and parametrized over the knot vector $\Uvec^{(0)}=\{0,0,0,0,1/6,1/3,1/2,2/3,5/6,1,1,1,1\}$ showing the asymmetry at the boundaries.
}
\label{fig:AsymmetricBasis}
\end{figure}

In this work, we employed clamped knot vectors (see Eq.~(\ref{eq:KnotVector})) where the first and last knots occur with a multiplicity of $p+1$.
This is required since the B-spline curve must pass through the first and last control point for $u=0$ and $u=1$, respectively, to allow for merging separate B-spline curves together.\cite{tai2003a}
As a result, clamping leads to an asymmetry of the $p$ first and last B-splines compared to the inner ones being completely symmetrical (see Fig.~\ref{fig:AsymmetricBasis}).
Accordingly, the data points lying at the boundary, associated with the parameters $u_0\leq \bar{u}_g<u_{2p-1}$ and $u_{n-p+2}< \bar{u}_g \leq u_{n+p+1}$, are represented differently by the curve.

\subsubsection{Matrix representation}

To conduct the penalized least-squares fitting procedure, Eq.\,\eqref{eq:PenalizedBSplineCurveApproximation} is written in matrix representation
\begin{align}
\left(\Nmat^\transp \Nmat + \lambda\, \Dmat^{(\kappa)\transp} \Dmat^{(\kappa)}\right) \Pmat = \Nmat^\transp\Qmat
\end{align}
with the $(m+1)\times(n+1)$ B-spline matrix
\begin{align}
\Nmat =
\begin{pmatrix}
\Nbas_{0,p}^{\Uvec}(\bar{u}_{0}) & \hdots & \Nbas_{n,p}^{\Uvec}(\bar{u}_{0})\\
\vdots & \ddots & \vdots\\
\Nbas_{0,p}^{\Uvec}(\bar{u}_{m}) & \hdots & \Nbas_{n,p}^{\Uvec}(\bar{u}_{m})\\
\end{pmatrix},
\end{align}

the
$(m+1)\times N$  and
$(n+1)\times N$ matrix
\begin{align}
\Pmat=
\begin{pmatrix}
\Pvec_{0}^\transp\\
\vdots\\
\Pvec_{n}^\transp
\end{pmatrix}
\quad
\text{and}
\quad
\Qmat=
\begin{pmatrix}
\Qvec_{0}^\transp\\
\vdots\\
\Qvec_{m}^\transp
\end{pmatrix}
,
\end{align}
respectively, and the
$(n+1-\kappa)\times(n+1)$ difference matrix
\begin{align}
\Dmat^{(\kappa)} =
\begin{pmatrix}
D_{0,0}^{(\kappa)} & \hdots & D_{0,n}^{(\kappa)}\\
\vdots & \ddots & \vdots\\
D_{n-\kappa,0}^{(\kappa)} & \hdots & D_{n-\kappa,n}^{(\kappa)}\\
\end{pmatrix},
\end{align}
resulting from the recursively defined difference operator $\Delta^{\kappa} = \Delta(\Delta^{\kappa-1})$.
For equally spaced knots, the $\kappa$-th order finite difference matrix elements $D_{i,j}^{(\kappa)}$ obtained from the recursively defined difference operator 
 can be calculated by
\begin{align}
D_{i,j}^{(\kappa)} &=
\begin{cases}
D_{i+1,j}^{\kappa-1} - D_{i,j}^{\kappa-1}, & k>1 \\
\delta_{i+1,j} - \delta_{i,j}, & k=1 
\end{cases}
\end{align}

with $\delta_{i,j}$ being the Kronecker delta.

\section*{References}

\providecommand{\refin}[1]{\\ \textbf{Referenced in:} #1}

\end{document}